\documentclass[12pt]{article}

\author{Yu.~M.~Zinoviev
       \thanks{E-mail address: Yurii.Zinoviev@ihep.ru} \\
        {\it Institute for High Energy Physics} \\
        {\it Protvino, Moscow Region, 142280, Russia}}
\title{Note on antisymmetric spin-tensors}

\date{}

\begin{document}

\maketitle

\begin{abstract}
It was known for a long time that in $d=4$ dimensions it is impossible
to construct the Lagrangian for antisymmetric second rank spin-tensor
that will be invariant under the gauge transformations with
unconstrained spin-vector parameter. But recently a paper \cite{BKR09}
appeared where gauge invariant Lagrangians for antisymmetric
spin-tensors of arbitrary rank $n$ in $d > 2n$ were constructed
using powerful BRST approach. To clarify apparent contradiction, in
this note we carry a direct independent analysis of the most general
first order Lagrangian for the massless antisymmetric spin-tensor of
second rank. Our analysis shows that gauge invariant Lagrangian does
exist but in $d \ge 5$ dimensions only, while in $d=4$ this Lagrangian
becomes identically zero. As a byproduct, we obtain a very simple and
convenient form of this massless Lagrangian that makes deformation to
$AdS$ space and/or massive case a simple task as we explicitly show
here. Moreover, this simple form admits natural and straightforward
generalization on the case of massive antisymmetric spin-tensors of
rank $n$ for $d > 2n$.
\end{abstract}

\thispagestyle{empty}
\newpage
\setcounter{page}{1}

\section*{Introduction}

It was known for a long time that in $d=4$ dimensions it is impossible
to construct the Lagrangian for antisymmetric second rank spin-tensor
that will be invariant under the gauge transformations with
unconstrained spin-vector parameter \cite{Tow80,DW80,DTS81} (see also
more recent paper \cite{NN04}). But recently a paper \cite{BKR09}
appeared where gauge invariant Lagrangians for antisymmetric
spin-tensors of arbitrary rank $n$ in $d > 2n$ were constructed
using powerful BRST approach 
\cite{BK05,BKRT06,BKL06,BKR07,MR07,BKT07,BKR08}. The aim of this short
note is to clarify an apparent contradiction between these new and
previous results.

The paper structured as follows. In Section 1, as a warming up
exercise, we consider simplest and very well known case of spin-vector
field $\Psi_\mu$. Such field can be considered as a simplest example
of completely antisymmetric spin-tensors and indeed this case gives
some useful hints. Then in Section 2 we consider the most general
first order Lagrangian for massless antisymmetric spin-tensor
$\Psi_{\mu\nu}$ in flat Minkowski space and show that the Lagrangian
invariant under the gauge transformations $\delta \Psi_{\mu\nu} =
\partial_\mu \xi_\nu - \partial_\nu \xi_\mu$ with unconstrained
spin-vector parameter $\xi_\mu$ does exist, but in $d \ge 5$
dimensions only, while it becomes identically zero in $d=4$. As a
byproduct we obtain a very simple and convenient form of this gauge
invariant Lagrangian that makes deformation into $AdS$ space a simple
task as we explicitly show here.

In Section 3 we consider gauge invariant description of massive
antisymmetric spin-tensor using a minimalistic approach as in
\cite{KZ97,Zin01,Met06,Zin08b,Zin08c,BG08}. Due to reducibility of
gauge transformations it turns out enough to  have two fields ---
spin-tensor $\Psi_{\mu\nu}$ and spin-vector $\Phi_\mu$ only. At last,
in Section 4 we show that these results admit a natural and
straightforward generalization on the case of spin-tensors with
arbitrary rank $n$ in $d \ge 2n+1$. Again, due to reducibility of
gauge transformations, to obtain gauge invariant description of
massive field it is enough to introduce two spin-tensors with
rank $n$ and $n-1$ only.

\section{Warming up exercise}

As a warming up exercise let us take an example of spin-vector
$\Psi_\mu$ (any way it can be considered as a simplest case of
completely antisymmetric spin-tensors). The most general first order
Lagrangian for massless field in flat Minkowski space can be written
as follows:
\begin{equation}
\frac{1}{i} {\cal L}_0 = \bar{\Psi}^\mu \hat{\partial} \Psi_\mu + a_1
( (\bar{\Psi} \partial) (\gamma \Psi) + (\bar{\Psi}\gamma) (\partial
\Psi)) + a_2 (\bar{\Psi}\gamma) \hat{\partial} (\gamma\Psi)
\end{equation}
Now if we require that this Lagrangian be invariant under the gauge
transformations $\delta \Psi_\mu = \partial_\mu \xi$ with spinor
parameter, we immediately obtain:
$$
\partial^\mu \frac{\delta {\cal L}_0}{\delta \bar{\Psi}^\mu} = 0 \quad
\Rightarrow \quad a_1 = - 1, \quad a_2 = 1
$$
First signal that something happens then $d=2$ comes from the
contraction of equation with $\gamma$-matrix:
$$
\gamma^\mu  \frac{\delta {\cal L}_0}{\delta \bar{\Psi}^\mu} = i (d-2)
[ \hat{\partial} (\gamma \Psi) - (\partial \Psi) ]
$$
Noting that in $d=2$ the last term in the Lagrangian containing three
$\gamma$-matrices is not independent, let us rewrite the Lagrangian in
terms of completely antisymmetric products of $\gamma$-s. Using the
following relation:
$$
\gamma^\mu \gamma^\nu \gamma^\alpha = g^{\mu\nu} \gamma^\alpha -
g^{\mu\alpha} \gamma^\nu + g^{\nu\alpha} \gamma^\mu +
\Gamma^{\mu\nu\alpha}
$$
where $\Gamma^{\mu\nu\alpha} = \gamma^{[\mu} \gamma^\nu
\gamma^{\alpha]}$, the most general Lagrangian can be written as
follows:
\begin{eqnarray}
\frac{1}{i} {\cal L}_0 &=& (1-a_2) \bar{\Psi}^\mu \hat{\partial}
\Psi_\mu + (a_1+a_2) ((\bar{\Psi}\partial) (\gamma \Psi) +
(\bar{\Psi} \gamma) (\partial \Psi)) + a_2 \bar{\Psi}_\mu
\Gamma^{\mu\nu\alpha} \partial_\nu \Psi_\alpha = \nonumber \\
 &=&  \bar{\Psi}_\mu \Gamma^{\mu\nu\alpha} \partial_\nu \Psi_\alpha
\end{eqnarray}
where the last line corresponds to the same values of parameters when
the Lagrangian is gauge invariant. In this form gauge invariance of
the Lagrangian is evident as well as the fact that it is identically
zero in $d=2$.

\section{Massless case}

In this Section we consider massless antisymmetric spin-tensor
$\Psi_{\mu\nu}$. The most general first order Lagrangian can be
written in the following form:
\begin{eqnarray}
\frac{1}{i} {\cal L}_0 &=& \bar{\Psi}^{\mu\nu} \hat{\partial}
\Psi_{\mu\nu} + a_1 ( (\bar{\Psi} \partial)^\mu (\gamma \Psi)_\mu +
(\bar{\Psi} \gamma)^\mu (\partial \Psi)_\mu ) + a_2 
(\bar{\Psi} \gamma)^\mu \hat{\partial} (\gamma \Psi)_\mu + \nonumber
\\
 && +  a_3 ( (\bar{\Psi} \partial \gamma) (\gamma \gamma \Psi) -
(\bar{\Psi} \gamma \gamma) (\partial \gamma \Psi) ) + a_4
(\bar{\Psi} \gamma \gamma) \hat{\partial} (\gamma \gamma \Psi)
\end{eqnarray}
where $(\gamma \Psi)_\nu = \gamma^\mu \Psi_{\mu\nu}$,
$(\bar{\Psi} \gamma)^\nu = \bar{\Psi}^{\mu\nu} \gamma_\mu$ and so on.
Now if we require that this Lagrangian be invariant under the gauge
transformations $\delta \Psi_{\mu\nu} = \partial_\mu \xi_\nu -
\partial_\nu \xi_\mu$ with unconstrained spin-vector parameter
$\xi_\mu$, we get:
$$
\partial^\mu \frac{\delta {\cal L}_0}{\delta \bar{\Psi}^{\mu\nu}} = 0
\quad \Rightarrow \quad a_1 = - 2, \quad a_2 = 2, \quad a_3 = - 1,
\quad a_4 = - \frac{1}{2}
$$
Again a first signal that something happens in $d=4$ comes from the
contraction with $\gamma$-matrix:
$$
\gamma^\mu \frac{\delta {\cal L}_0}{\delta \bar{\Psi}^{\mu\nu}} =
\frac{i(d-4)}{2} [ - 2 (\partial \Psi)_\nu + 2 \hat{\partial} 
(\gamma \Psi)_\nu + 2 \gamma_\nu (\partial \gamma \Psi) + \partial_\nu
(\gamma \gamma \Psi) - \gamma_\nu \hat{\partial} (\gamma \gamma \Psi)]
$$
Noting that in $d=4$ the last term in the Lagrangian containing five
$\gamma$-s is not independent, we proceed rewriting the Lagrangian
in terms of completely antisymmetric products. Using the relation:
\begin{eqnarray*}
\Gamma^{\mu\nu} \gamma^\alpha \Gamma^{\beta\gamma} &=& -
(g^{\mu\beta} g^{\nu\gamma} - g^{\mu\gamma} g^{\nu\beta})
\gamma^\alpha + (- g^{\mu\alpha} g^{\nu\beta} \gamma^\gamma +
g^{\mu\alpha} g^{\nu\gamma} \gamma^\beta + g^{\nu\alpha} g^{\mu\beta}
\gamma^\gamma - g^{\nu\alpha} g^{\mu\gamma} \gamma^\beta) + \\
 && + (g^{\nu\gamma} g^{\alpha\beta} \gamma^\mu - g^{\mu\gamma}
g^{\alpha\beta} \gamma^\nu - g^{\nu\beta} g^{\alpha\gamma} \gamma^\mu
+ g^{\mu\beta} g^{\alpha\gamma} \gamma^\nu ) + (g^{\nu\alpha}
\Gamma^{\mu\beta\gamma} - g^{\mu\alpha} \Gamma^{\nu\beta\gamma}) + \\
 && + (g^{\alpha\beta} \Gamma^{\mu\nu\gamma} - g^{\alpha\gamma}
\Gamma^{\mu\nu\beta}) + (g^{\mu\beta} \Gamma^{\nu\alpha\gamma} -
g^{\nu\beta} \Gamma^{\mu\alpha\gamma} - g^{\mu\gamma}
\Gamma^{\nu\alpha\beta} + g^{\nu\gamma} \Gamma^{\mu\alpha\beta}) + 
\Gamma^{\mu\nu\alpha\beta\gamma} 
\end{eqnarray*}
we obtain for the most general Lagrangian:
\begin{eqnarray*}
\frac{1}{i} {\cal L}_0 &=& (1-a_2-2a_4) \bar{\Psi}^{\mu\nu}
\hat{\partial} \Psi_{\mu\nu} + (a_1+a_2-2a_3+4a_4) ( (\bar{\Psi}
\partial)^\mu (\gamma \Psi)_\mu + (\bar{\Psi} \gamma)^\mu 
(\partial \Psi)_\mu) + \\
 && + (a_2+4a_4) \bar{\Psi}_{\mu\nu} \Gamma^{\nu\alpha\beta}
\Psi_{\mu\beta} + (a_3-2a_4) ( (\bar{\Psi} \partial)_\mu
\Gamma^{\mu\nu\alpha} \Psi_{\nu\alpha} - \bar{\Psi}_{\mu\nu}
\Gamma^{\mu\nu\alpha}  (\partial \Psi)_\alpha)  + \\
 && + a_4 \bar{\Psi}_{\mu\nu} \Gamma^{\mu\nu\alpha\beta\gamma}
\partial_\alpha \Psi_{\beta\gamma} = \\
 &=& - \frac{1}{2} \bar{\Psi}_{\mu\nu}
\Gamma^{\mu\nu\alpha\beta\gamma} \partial_\alpha \Psi_{\beta\gamma}
\end{eqnarray*}
where the last line corresponds to gauge invariant case. Again in this
form the gauge invariance of the Lagrangian is evident as well as the
fact that it is identically zero in $d=4$.

The last form of the Lagrangian is very simple and convenient. In
particular, it makes deformation into $AdS$ space a simple task.
Indeed, let us consider the Lagrangian:
\begin{equation}
{\cal L}_0 = \frac{i}{2} \bar{\Psi}_{\mu\nu}
\Gamma^{\mu\nu\alpha\beta\gamma} D_\alpha \Psi_{\beta\gamma}
\end{equation}
where ordinary partial derivative is replaced by $AdS$ covariant one.
Note that the Lagrangian is completely antisymmetric in all vector
indices, so that covariant derivative effectively acts on the spinor
index only, e.g.:
$$
 [ D_\mu, D_\nu ] \xi_\alpha = - \frac{\kappa}{2} \Gamma_{\mu\nu}
\xi_\alpha, \qquad \kappa = \frac{2 \Lambda}{(d-1)(d-2)}
$$
Due to non-commutativity of covariant derivatives this Lagrangian is
not invariant under the covariantized gauge transformations $\delta
\Psi_{\mu\nu} = D_\mu \xi_\nu - D_\nu \xi_\mu$ any more:
$$
\delta_0 {\cal L}_0 = \frac{i\kappa}{4} (d-3)(d-4) \bar{\Psi}_{\mu\nu}
\Gamma^{\mu\nu\alpha} \xi_\alpha
$$
To compensate this non-invariance we add mass-like term to the
Lagrangian as well as corresponding correction to gauge
transformations:
\begin{equation}
{\cal L}_1 = \frac{M}{2} \bar{\Psi}_{\mu\nu}
\Gamma^{\mu\nu\alpha\beta} \Psi_{\alpha\beta}, \qquad \delta_1
\Psi_{\mu\nu} = i \alpha_0 (\gamma_\mu \xi_\nu - \gamma_\nu \xi_\mu)
\end{equation}
Simple calculations show that the gauge invariance of the Lagrangian
can indeed be restored provided:
$$
\alpha_0 = - \frac{M}{(d-4)}, \qquad
M^2 = - \frac{(d-4)^2}{4} \kappa
$$
where the last relation clearly shows that the procedure works in
$AdS$ space ($\kappa < 0$) only. Thus we obtain the following
Lagrangian and gauge transformations:
\begin{eqnarray}
{\cal L}_0 &=& \frac{i}{2} \bar{\Psi}_{\mu\nu}
\Gamma^{\mu\nu\alpha\beta\gamma} D_\alpha \Psi_{\beta\gamma} +
\frac{M}{2} \bar{\Psi}_{\mu\nu}
\Gamma^{\mu\nu\alpha\beta} \Psi_{\alpha\beta} \\
\delta \Psi_{\mu\nu} &=& D_\mu \xi_\nu - D_\nu \xi_\mu - 
\frac{iM}{(d-4)} (\gamma_\mu \xi_\nu - \gamma_\nu \xi_\mu)
\end{eqnarray}
describing massless antisymmetric second rank tensor in $AdS_d$ space
with $d \ge 5$.

\section{Massive case}

In this Section we consider gauge invariant description of massive
antisymmetric second rank tensor in arbitrary $(A)dS_d$ space with $d
\ge 5$. Spin-tensor $\Psi_{\mu\nu}$ has gauge transformations with
spin-vector parameter, so for gauge invariant description of massive
tensor we have introduce additional Goldstone field $\Phi_\mu$. This
field has its own gauge transformations with spinor parameter, but due
to reducibility of gauge transformations for spin-tensor
$\Psi_{\mu\nu}$ it is not necessary to introduce any other additional
fields. We have already seen that completely antisymmetric products of
$\gamma$ matrices plays an essential role, so we will look for the
Lagrangian in the form:
\begin{eqnarray}
{\cal L} &=& \frac{i}{2} \bar{\Psi}_{\mu\nu}
\Gamma^{\mu\nu\alpha\beta\gamma} D_\alpha \Psi_{\beta\gamma} + i
\bar{\Phi}_\mu \Gamma^{\mu\nu\alpha} D_\nu \Phi_\alpha + \nonumber \\
 && + \frac{M}{2} \bar{\Psi}_{\mu\nu} \Gamma^{\mu\nu\alpha\beta}
\Psi_{\alpha\beta} + ia_1 ( \bar{\Psi}_{\mu\nu} \Gamma^{\mu\nu\alpha}
\Phi_\alpha + \bar{\Phi}_\mu \Gamma^{\mu\nu\alpha} \Psi_{\nu\alpha})
+ a_2 \bar{\Phi}_\mu \Gamma^{\mu\nu} \Phi_\nu
\end{eqnarray}
At the same time we will use the following ansatz for the gauge
transformations:
\begin{eqnarray}
\delta \Psi_{\mu\nu} &=& D_\mu \xi_\nu - D_\nu \xi_\mu + i \alpha_0
(\gamma_\mu \xi_\nu - \gamma_\nu \xi_\mu) + \alpha_1 \Gamma_{\mu\nu}
\eta \nonumber \\
\delta \Phi_\mu &=& D_\mu \eta + \alpha_2 \xi_\mu + i \alpha_3
\gamma_\mu \eta
\end{eqnarray}
First of all we calculate all variations with one derivative and
require their cancellation. It allows us to express all parameters in
gauge transformations in terms of the ones in the Lagrangian:
$$
\alpha_0 = - \frac{M}{(d-4)}, \quad 
\alpha_1 = \frac{2 a_1}{(d-3)(d-4)},
\quad \alpha_2 = - 2 a_1, \quad \alpha_3 = \frac{a_2}{d-2}
$$
Now we calculate all variations without derivatives taking into
account contributions of kinetic terms due to non-commutativity of
covariant derivatives. Their cancellation gives:
$$
M^2 = \frac{2(d-4)}{(d-3)} a_1{}^2 - \frac{(d-4)^2}{16} \kappa,
\quad a_2 = - \frac{(d-2)}{(d-4)} M
$$
Working with gauge invariant description of massive fields it is
natural to define massless limit as a one where (all) Goldstone
field(s) decouple from the main gauge one. For the case at hands we
see that it is the limit $a_1 \to 0$ that corresponds to the massless
one. As for the concrete mass normalization, we will use convention
"mass is what would be mass in flat Minkowski space". Using the
following relation:
$$
\bar{\Psi}^{\mu\nu} \Psi_{\mu\nu} - 2 (\bar{\Psi} \gamma)^\mu
(\gamma \Psi)_\mu - \frac{1}{2} (\bar{\Psi} \gamma \gamma) 
(\gamma \gamma \Psi) = - \frac{1}{2} \bar{\Psi}_{\mu\nu}
\Gamma^{\mu\nu\alpha\beta} \Psi_{\alpha\beta}
$$
we obtain:
$$
a_1{}^2 = \frac{(d-3)}{2(d-4)} m^2, \qquad M^2 = m^2 -
\frac{(d-4)^2}{4} \kappa
$$
Collecting all results together we obtain the following Lagrangian:
\begin{eqnarray}
{\cal L} &=& \frac{i}{2} \bar{\Psi}_{\mu\nu}
\Gamma^{\mu\nu\alpha\beta\gamma} D_\alpha \Psi_{\beta\gamma} + i
\bar{\Phi}_\mu \Gamma^{\mu\nu\alpha} D_\nu \Phi_\alpha + 
\frac{M}{2} \bar{\Psi}_{\mu\nu} \Gamma^{\mu\nu\alpha\beta}
\Psi_{\alpha\beta} + \nonumber \\
 && +  i \sqrt{\frac{(d-3)}{2(d-4)}} m 
( \bar{\Psi}_{\mu\nu} \Gamma^{\mu\nu\alpha}
\Phi_\alpha + \bar{\Phi}_\mu \Gamma^{\mu\nu\alpha} \Psi_{\nu\alpha})
- \frac{(d-2)}{(d-4)} M \bar{\Phi}_\mu \Gamma^{\mu\nu} \Phi_\nu
\end{eqnarray}
invariant under the following gauge transformations:
\begin{eqnarray}
\delta \Psi_{\mu\nu} &=& D_\mu \xi_\nu - D_\nu \xi_\mu - 
\frac{2iM}{(d-4)} (\gamma_\mu \xi_\nu - \gamma_\nu \xi_\mu) +
\sqrt{\frac{2}{(d-3)(d-4)}} \frac{m}{(d-4)} \Gamma_{\mu\nu} \eta
\nonumber \\
\delta \Phi_\mu &=& D_\mu \eta - \sqrt{\frac{2(d-3)}{(d-4)}} m \xi_\mu
- i \frac{M}{(d-4)} \gamma_\mu \eta
\end{eqnarray}
where $M^2 = m^2 - \frac{(d-4)^2}{4} \kappa$. Note that this time our
construction works in $dS$ space ($\kappa > 0$) as well but for $m^2
\ge \frac{(d-4)^2}{4} \kappa$ only. At the boundary of this region (as
it is often to be the case for fermionic fields) the Lagrangian and
gauge transformations become much simpler:
\begin{eqnarray}
{\cal L} &=& \frac{i}{2} \bar{\Psi}_{\mu\nu}
\Gamma^{\mu\nu\alpha\beta\gamma} D_\alpha \Psi_{\beta\gamma} + i
\bar{\Phi}_\mu \Gamma^{\mu\nu\alpha} D_\nu \Phi_\alpha + \nonumber  \\
 && +  i \sqrt{\frac{(d-3)}{2(d-4)}} m 
( \bar{\Psi}_{\mu\nu} \Gamma^{\mu\nu\alpha}
\Phi_\alpha + \bar{\Phi}_\mu \Gamma^{\mu\nu\alpha} \Psi_{\nu\alpha})
\end{eqnarray}
\begin{eqnarray}
\delta \Psi_{\mu\nu} &=& D_\mu \xi_\nu - D_\nu \xi_\mu  +
\sqrt{\frac{2}{(d-3)(d-4)}} \frac{m}{(d-4)} \Gamma_{\mu\nu} \eta
\nonumber \\
\delta \Phi_\mu &=& D_\mu \eta - \sqrt{\frac{2(d-3)}{(d-4)}} m \xi_\mu
\end{eqnarray}

\section{Arbitrary rank}

In this Section we consider generalization of the above results to the
case of completely antisymmetric spin-tensors of arbitrary rank. The
higher the rank the more and more complicated become the calculations,
thus we will not consider the most general possibilities here.
Instead, simply assuming that completely antisymmetric products of
$\gamma$ matrices do the job, we will try to construct gauge invariant
description of massive spin-tensor of rank $n$ in $(A)dS_d$ space with
$d \ge 2n+1$. Again due to reducibility of gauge transformations we
will need two spin-tensors with ranks $n$ and $n-1$ only: 
$\Psi_{\mu_1 \dots \mu_n}$ and $\Phi_{\mu_1 \dots \mu_{n-1}}$. By
analogy with $n=2$ case we will look for the Lagrangian in the
following form:
\begin{eqnarray}
{\cal L} &=& \frac{i}{n!} \bar{\Psi}_{\mu_1 \dots \mu_n} 
\Gamma^{\mu_1 \dots \mu_n\alpha\nu_1 \dots \nu_n} D_\alpha
\Psi_{\nu_1 \dots \nu_n} + \frac{i}{(n-1)!} 
\bar{\Phi}_{\mu_1 \dots \mu_{n-1}} 
\Gamma^{\mu_1 \dots \mu_{n-1}\alpha\nu_1 \dots \nu_{n-1}} D_\alpha
\Phi_{\nu_1 \dots \nu_{n-1}} + \nonumber \\
 && + \frac{M}{n!} \bar{\Psi}_{\mu_1 \dots \mu_n}
\Gamma^{\mu_1 \dots \mu_n \nu_1 \dots \nu_n}
\Psi_{\nu_1 \dots \nu_n} + \frac{a_2}{(n-1)!}
\bar{\Phi}_{\mu_1 \dots \mu_{n-1}} 
\Gamma^{\mu_1 \dots \mu_{n-1}\nu_1 \dots \nu_{n-1}}
\Phi_{\nu_1 \dots \nu_{n-1}} + \nonumber \\
 && + \frac{i a_1}{(n-1)!} ( \bar{\Psi}_{\mu_1 \dots \mu_n}
\Gamma^{\mu_1 \dots \mu_n\nu_1 \dots \nu_{n-1}}
\Phi_{\nu_1 \dots \nu_{n-1}} +  \bar{\Phi}_{\mu_1 \dots \mu_{n-1}} 
\Gamma^{\mu_1 \dots \mu_{n-1}\nu_1 \dots \nu_n}
\Psi_{\nu_1 \dots \nu_n} )
\end{eqnarray}
Similarly, we introduce the following ansatz for gauge
transformations:
\begin{eqnarray}
\delta \Psi_{\mu_1 \dots \mu_n} &=& D_{[\mu_1} 
\xi_{\mu_2 \dots \mu_n]} + i \alpha_0 \gamma_{[\mu_1}
\xi_{\mu_2 \dots \mu_n]} + \alpha_1 \Gamma_{[\mu_1\mu_2}
\eta_{\mu_3 \dots \mu_n]} \nonumber \\
\delta \Phi_{\mu_1 \dots \mu_{n-1}} &=& D_{[\mu_1}
\eta_{\mu_2 \dots \mu_{n-1}]} + \alpha_2 \xi_{\mu_1 \dots \mu_{n-1}} +
i \alpha_3 \gamma_{[\mu_1} \eta_{\mu_2 \dots \mu_{n-1}]}
\end{eqnarray}
where brackets denote antisymmetrization with weight one. Again we
begin with the calculation of all variations with one derivative and
require their cancellation. It allows us to express all parameters in
gauge transformations in terms of the one in the Lagrangian:
$$
\alpha_0 = (-1)^{n-1} \frac{M}{(d-2n)}, \qquad
\alpha_1 = \frac{2 a_1}{(d-2n)(d-2n+1)}
$$
$$
\alpha_2 = - n a_1, \qquad
\alpha_3 = (-1)^n \frac{a_2}{(d-2n+2)}
$$
Then we calculate all variations without derivatives taking into
account contributions of kinetic terms due to non-commutativity of
covariant derivatives. And indeed all of them cancel provided:
$$
M^2 = \frac{n(d-2n)}{(d-2n+1)} a_1{}^2 - \frac{(d-2n)^2}{4}
\kappa, \qquad a_2 = - \frac{(d-2n+2)}{(d-2n)} M
$$
As in the previous case, we see that it is the limit $a_1 \to 0$
corresponds to the massless limit. In this, our convention on mass
normalization gives:
$$
a_1{}^2 = \frac{(d-2n+1)}{n(d-2n)} m^2, \qquad
M^2 = m^2 - \frac{(d-2n)^2}{4} \kappa
$$
Again the whole construction works in $dS$ space ($\kappa > 0$) as
well but for $m^2 \ge \frac{(d-2n)^2}{4} \kappa$ only.

\section*{Conclusion}

Thus really there is no contradiction between new and previous results
on antisymmetric second rank spin-tensor. The resolution turns out to
be simple (though may be unexpected): the gauge invariant Lagrangian
does exists in $d \ge 5$, but in $d=4$ it becomes zero thus explaining
why all previous attempts were unsuccessful. The simple completely
antisymmetric on all vector indices form of the Lagrangian obtained,
turns out to be very convenient in practical calculations such as
deformation into $AdS$ space or introduction of mass. Thus it seems
worthwhile to reconsider the problem of possible interactions for such
fields.


\begin{thebibliography}{10}

\bibitem{Tow80}
P.~K. Townsend
{\it "Gauge Invariance for Spin 1/2",}
Phys. Lett. {\bf B90} (1980) 275.

\bibitem{DW80}
S.~Deser, E.~Witten
{\it "Dynamical Properties of Antisyymetric Tensor Fields",}
Nucl. Phys. {\bf B178} (1980) 491.

\bibitem{DTS81}
S.~Deser, P.~K. Townsend, W.~Siegel
{\it "Higher Rank Representations of Lower Spin",}
Nucl. Phys. {\bf B184} (1981) 333.

\bibitem{NN04}
J.~Niederle, A.~G. Nikitin
{\it "Relativistic wave equations for interacting massive particles
with arbitrary half-intreger spins",}
Phys. Rev. {\bf D64} (2001) 125013, arXiv:hep-th/0412213.

\bibitem{BKR09}
I.~L. Buchbinder, V.~A. Krykhtin, L.~L. Ryskina
{\it "Lagrangian formulation of massive fermionic totally
antisymmetric tensor field theory in $AdS_d$ space",}
arXiv:0902.1471.

\bibitem{BK05}
I.~L. Buchbinder, V.~A. Krykhtin
{\it "Gauge invariant Lagrangian construction for massive bosonic
higher spin fields in D dimensions",}
Nucl. Phys. {\bf B727} (2005) 537, arXiv:hep-th/0505092.

\bibitem{BKRT06}
I.~L. Buchbinder, V.~A. Krykhtin, L~.L. Ryskina, H.~Takata
{\it "Gauge invariant Lagrangian construction for massive higher spin
fermionic fields",}
Phys. Lett. {\bf B641} (2006) 386, arXiv:hep-th/0603212.

\bibitem{BKL06}
I.~L. Buchbinder, V.~A. Krykhtin, P.~M. Lavrov
{\it "Gauge invariant Lagrangian formulation of higher spin massive
bosonic field theory in AdS space",}
Nucl. Phys. {\bf B762} (2007) 344, arXiv:hep-th/0608005.

\bibitem{BKR07}
I.~L. Buchbinder, V.~A. Krykhtin, A.~A. Reshetnyak
{\it "BRST approach to Lagrangian construction for fermionic higher
spin fields in (A)dS space",}
Nucl. Phys. {\bf B787} (2007) 211, arXiv:hep-th/0703049.

\bibitem{MR07}
P.~Yu. Moshin, A.~A. Reshetnyak
{\it "BRST approach to Lagrangian formulation for mixed-symmetry
fermionic higher-spin fields",}
JHEP {\bf 10} (2007) 040, arXiv:0707.0386.

\bibitem{BKT07}
I.~L. Buchbinder, V.~A. Krykhtin, H.~Takata
{\it "Gauge invariant Lagrangian construction for massive bosonic
mixed symmetry higher spin fields",}
Phys. Lett. {\bf B656} (2007) 253, arXiv:0707.2181.

\bibitem{BKR08}
I.~L. Buchbinder, V.~A. Krykhtin, L.~L. Ryskina
{\it "BRST approach to Lagrangian formulation of bosonic totally
antisymmeric tensor fields in curved space",} arXiv:0810.3467.

\bibitem{KZ97}
S.~M. Klishevich, Yu.~M. Zinoviev
{\it "On electromagnetic interaction of massive spin-2 particle",}
Phys. Atom. Nucl. {\bf 61} (1998) 1527, arXiv:hep-th/9708150.

\bibitem{Zin01}
Yu.~M. Zinoviev
{\it "On Massive High Spin Particles in (A)dS",}
arXiv:hep-th/0108192.

\bibitem{Met06}
R.~R. Metsaev
{\it "Gauge invariant formulation of massive totally symmetric
fermionic fields in (A)dS space",}
Phys. Lett. {\bf B643} (2006) 205-212, arXiv:hep-th/0609029.

\bibitem{Zin08b}
Yu.~M. Zinoviev
{\it "Frame-like gauge invariant formulation for massive high spin
particles",}
Nucl. Phys. {\bf B808} (2009) 185, arXiv:0808.1778.

\bibitem{Zin08c}
Yu.~M. Zinoviev
{\it "Towards frame-like gauge invariant formulation for massive mixed
symmetry bosonic fields",}
Nucl. Phys. {\bf B812} (2009) 46, arXiv:0809.3287.

\bibitem{BG08}
I.~L. Buchbinder, A.~V. Galajinsky
{\it "Quartet unconstrained formulation for massive higher spin
fields",}
JHEP {\bf 0811} (2008) 081, arXiv:0810.2852.

\end{thebibliography}
\end{document}